\documentclass[fleqn,usenatbib]{mnras}

\usepackage{newtxtext,newtxmath}

\usepackage[T1]{fontenc}
\usepackage{ae,aecompl}

\usepackage{graphicx}   
\usepackage{amsmath}    
\usepackage{amssymb}    

\usepackage{breakurl}

\newcommand{\sunmass}{~M$_{\odot}$~}
\newcommand{\kms}{~km s$^{-1}$~}
\newcommand{\TCRB}{T~CrB~}
\newcommand{\TCRBE}{T~CrB}
\newcommand{\dotM}{~M$_{\odot}$~yr$^{-1}$~}
\newcommand{\XMM}{{\it XMM-Newton~}}

\newcommand{\Swift}{{\it Swift~}}
\newcommand{\SwiftE}{{\it Swift}}

\newcommand{\NuSTAR}{{\it NuSTAR~}}

\newcommand{\Suzaku}{{\it Suzaku~}}
\newcommand{\SuzakuE}{{\it Suzaku}}
\newcommand{\xspec}{{\sc xspec~}}
\newcommand{\xspecE}{{\sc xspec}}

\title[X-rays from \TCRBE]
{\XMM observations of the symbiotic recurrent nova \TCRBE:
evolution of X-ray emission during the active phase}

\author[S.A.Zhekov and T.V.Tomov]{Svetozar A. Zhekov$^1$\thanks{
E-mail: szhekov@astro.bas.bg; toma.tomov@astri.umk.pl.} 
and Toma V. Tomov$^2$\\
$^1$Institute of Astronomy and National Astronomical Observatory
(Bulgarian Academy of Sciences),\\
72 Tsarigradsko Chaussee Blvd., Sofia 1784, Bulgaria\\
$^2$Centre for Astronomy, Faculty of Physics, Astronomy and
Informatics, Nicolaus Copernicus University, \\
Grudziadzka 5, 87-100 Torun, Poland
}

\date{}

\pubyear{2019}

\begin{document}
\label{firstpage}
\pagerange{\pageref{firstpage}--\pageref{lastpage}}
\maketitle

\begin{abstract}
We present an analysis of the \XMM observations of the symbiotic
recurrent nova \TCRBE, obtained during its active phase which started
in 2014 - 2015. 
The \XMM spectra of \TCRB have two prominent components: a
soft one (0.2 - 0.6 keV), well represented by black-body emission, 
and a heavily absorbed hard component (2 - 10 keV), well matched by 
optically-thin plasma emission with high temperature 
(kT $\approx 8$ keV).
The \XMM observations reveal evolution of the X-ray emission from
\TCRB in its active phase. Namely, the soft component in its spectrum
is decreasing with time while the opposite is true for
the hard component. 
Comparison with data obtained in the quiescent phase 
shows that the soft component is typical {\it only} for the
active phase, while the hard component is present in both phases but 
it is considerably stronger in the quiescent phase.
Presence of stochastic variability (flickering) on time-scales of 
minutes and hours is confirmed both in X-rays and UV (UVM2 filter of 
the \XMM optical monitor). On the other hand, periodic variability of
6000-6500 s is found for the first time in the soft X-ray emission 
(0.2 - 0.6 keV) from \TCRBE.
We associate this periodic variability with the rotational period of
the white dwarf in this symbiotic binary.

\end{abstract}




\section{Introduction}
\TCRB (HD 143454) is a symbiotic recurrent nova with two nova-like
outbursts recorded by the modern astronomy in 1868 and 1946. 
It is a binary system with an orbital period of 227.57 days
\citep{fekel_00} at a Gaia distance of $806^{+33}_{-31}$ pc
\citep{bailer_jones_18}.
The cool star in this binary system is a red giant of a M4III spectral type
\citep{murset_99} that fills its Roche lobe which facilitates accretion
onto the more massive hot component, a white dwarf
(\citealt{bel_mik_98}; \citealt{stanishev_04}). 

The first X-ray detection of \TCRB was with the {\it Einstein}
observatory \citep{cordova_81}, followed by studies
with various X-ray observatories in the last two decades.
The X-ray emission of \TCRB is hard and highly absorbed
(\citealt{luna_08}; \citealt{kennea_09}; \citealt{luna_18}),
so, it is a typical member of the class $\delta$ of the X-ray sources
among symbiotic stars: the X-rays from these objects likely
originate from the boundary layer between an accretion disk and the
white dwarf \citep{luna_13}. Its X-ray emission shows strong
stochastic variability both in the soft ($< 10$ keV) and hard ($> 20$
keV) energy bands (e.g., \citealt{luna_08}; \citealt{ilkiewicz_16})
that resembles the characteristics of the flicker noise (flickering),
typical for accretion processes in astrophysical objects.

Recently, \citet{munari_16} reported that in 2015 \TCRB has entered a 
phase of unprecedented activity
(but the optical brightening started yet in 2014), 
a super-active phase as named by
these authors, that is similar to its state a few years before its
nova-like eruption in 1946. However, \citet{ilkiewicz_16} provided 
arguments that in the past \TCRB has experienced numerous active 
phases and this in 2015 is just one such an active 
phase\footnote{Throughout this text, the term {\it active} phase will 
be used to denote the activity of \TCRB which started in 
2014 - 2015}.
So, 2014-2015 
marked the start of new activities in \TCRB that 
manifest in optical, UV and X-rays and will provide valuable pieces
of information about the physics of this symbiotic recurrent nova.

In this paper, we report results from the \XMM observations of \TCRBE,
carried out during its current active phase.
In Section~\ref{sec:data}, we review the observational data.
In Section~\ref{sec:results}, we present results from analysis of the
X-ray emission of \TCRBE. In Section~\ref{sec:discussion}, we discuss 
our results and Section~\ref{sec:conclusions} presents our 
conclusions.

\begin{figure}
\begin{center}
\includegraphics[width=\columnwidth]{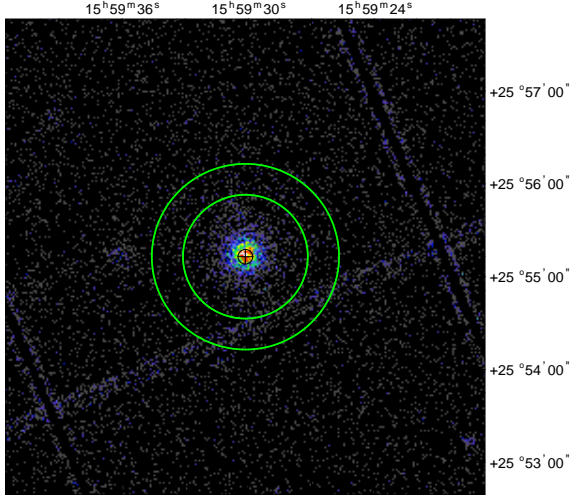}
\end{center}
\caption{The raw EPIC-pn image of \TCRB in the (0.2 - 10 keV) energy
band with the spectral extraction regions. The source spectrum was
extracted from the central circle, while the background spectrum was
extracted from adjacent annulus. The circled plus sign gives the
optical position of \TCRB (SIMBAD).
}
\label{fig:image}
\end{figure}

\section{Observations and data reduction}
\label{sec:data}

\TCRB was observed with \XMM on 2018 January 30 (Observation ID
0800420201; PI: S.Zhekov) with a nominal exposure of $\sim 28.0$ ks. 
The source was clearly 
detected in X-rays (see Fig.~\ref{fig:image}) and good quality data
were acquired with the European Photon Imaging Camera 
(EPIC) having one pn and two MOS detectors. 
The \XMM Optical Monitor (OM)\footnote{for EPIC and OM see \S~3.3 and 
3.5 in the \XMM Users Handbook,
\url{https://xmm-tools.cosmos.esa.int/external/xmm_user_support/documentation/uhb/}
}
telescope allows for obtaining optical/UV data on a given target 
simultaneously with its X-ray data. Thus, six UV exposures in UVM2
filter were obtained that provided good quality UV light curves (LC) 
of \TCRBE.

{\it X-rays.}
For the data reduction, we made use of the \XMM 
{\sc sas}\footnote{Science Analysis Software, 
\url{https://xmm-tools.cosmos.esa.int/external/xmm_user_support/documentation/sas\_usg/USG/}}
16.1.0 data analysis software.
The {\sc sas} pipeline processing scripts emproc and  epproc 
were executed to incorporate the most recent calibration files (as of 
2017 December 20). The data were then filtered for high X-ray 
background following the instructions in the {\sc sas} documentation. 
The corresponding {\sc sas} procedures were adopted to generate the
response matrix files and ancillary response files for 
each spectrum. The MOS spectrum in our analysis is the sum of the
spectra from the two MOS detectors. 
The extracted EPIC spectra (0.2 - 10 keV) of \TCRB had
$\sim 2848 $~source counts in the 20.1-ks pn effective exposure and 
$\sim 1538$~source counts in the 26.3-ks MOS effective exposure.
Also, we constructed the pn and MOS1,2 background-subtracted light 
curves of \TCRBE.

{\it UV.}
We used the pipeline background-subtracted light 
curves of \TCRB in the UVM2 filter (effective wavelength and width of 
2310 \AA ~and 480 \AA, respectively) in units of counts s$^{-1}$ which
can be presented also in magnitudes using the zeropoint for
the UVM2 filter. 

Similarly, the data reduction (in X-rays) was done for the 
archive observation of \TCRB carried out on 2017 Febuary 23 
(Observation ID 0793183601; Target of Opportunity) that provided 
EPIC spectra (0.2 - 10 keV) of \TCRB with
$\sim 9230 $~source counts in the 37.5-ks pn effective exposure and
$\sim 3634$~source counts in the 59.3-ks MOS effective exposure.
We note that these data were already discussed in \citet{luna_18}.

For the spectral analysis, we used 
version 12.9.1 of \xspec \citep{Arnaud96}.

\section{Results}
\label{sec:results}

\subsection{An overview of the X-ray emission}
\label{sec:xray_overview}
The \XMM observations of \TCRB during its current active phase
provide data of good quality that reveal some interesting global
features of the X-ray emission from this symbiotic system. 

In their analysis of the 2017 \XMM observation,
\citet{luna_18} pointed out that there were two strong components in
the (0.2 - 10 keV) X-ray spectrum of \TCRBE. Namely, a soft one that
dominates the spectrum in the 0.2 - 0.6 keV energy range (well matched
by black-body emission), and a highly absorbed  hard component at 
energies above 1.5 -  2 keV. It is interesting to note that the
soft component is a new feature in the X-ray emission from \TCRB that
appeared only in the current active phase and was not present before
that as illustrated by the \Suzaku spectra \citep{luna_08}.

\begin{figure*}
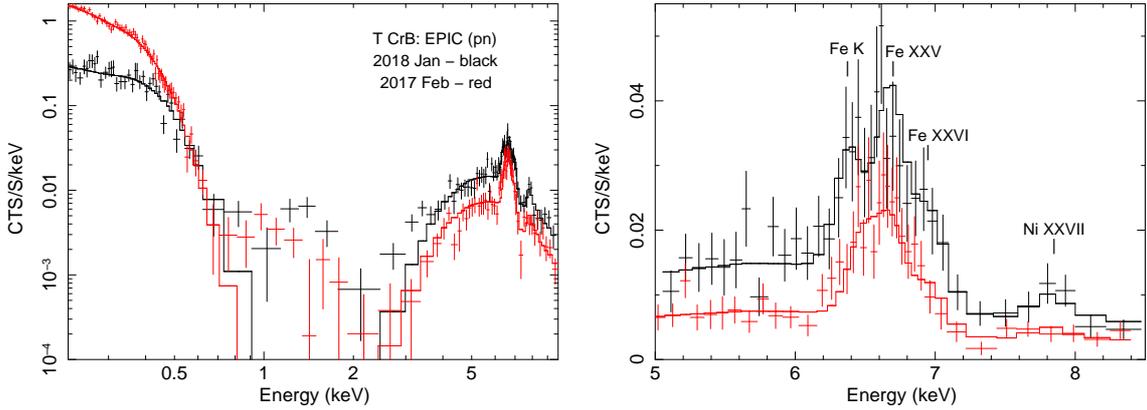

\begin{center}
\includegraphics[width=2.1in, height=3.0in, angle=-90]{fig2a.eps}
\includegraphics[width=2.1in, height=3.0in, angle=-90]{fig2b.eps}
\end{center}
\caption{ The background-subtracted pn spectra of \TCRB (left panel): 
the model fits are shown with the solid line (see text;
Section~\ref{sec:xray_spec}).  Emission 
lines of thermal origin are seen in the high-energy region 
(5 - 8 keV; right panel): marked are the K-shell fluorescent Fe lines
at $\sim 6.4$~keV (Fe K); the iron He-like triplet at $\sim 6.7$~keV
(Fe XXV); the Fe XXVI L$_\alpha$ at $\sim 6.97$~keV (Fe XXVI) and the
nickel He-like triplet at $\sim 7.8$~keV (Ni XXVII).
Spectra are re-binned to have a minimum of 30 counts per bin. 
}
\label{fig:pn_spectra}
\end{figure*}

As seen from Fig.~\ref{fig:pn_spectra} (left panel), these two 
components still dominate in the 2018 spectrum of \TCRB with some 
noticeable changes: the soft component is becoming weaker with time 
(2018 vs 2017) while the opposite is valid for the hard component. 
Thanks to the higher hard X-ray flux, some spectral-line features 
are better seen in the 2018 spectrum of \TCRB 
(Fig.~\ref{fig:pn_spectra}; right panel), which is a basic
illustration of the thermal origin of the hard component in its X-ray
emission.

On the other hand, it is worth mentioning that there is a weak X-ray
emission at `intermediate' energies (0.6 - 2 keV; see
Fig.~\ref{fig:pn_spectra}, left panel), detected at a $> 5\sigma$~
level in all the EPIC spectra (pn and MOS). So, this emission is not a
result from bad background subtraction and should be thus associated 
with \TCRBE. As seen from Fig.~\ref{fig:pn_spectra}, the
`intermediate' component correlates with the hard component in the
spectrum: its emission is stronger in 2018 than in 2017.  We will
return to this in Section \ref{sec:xray_spec}.

\begin{figure*}
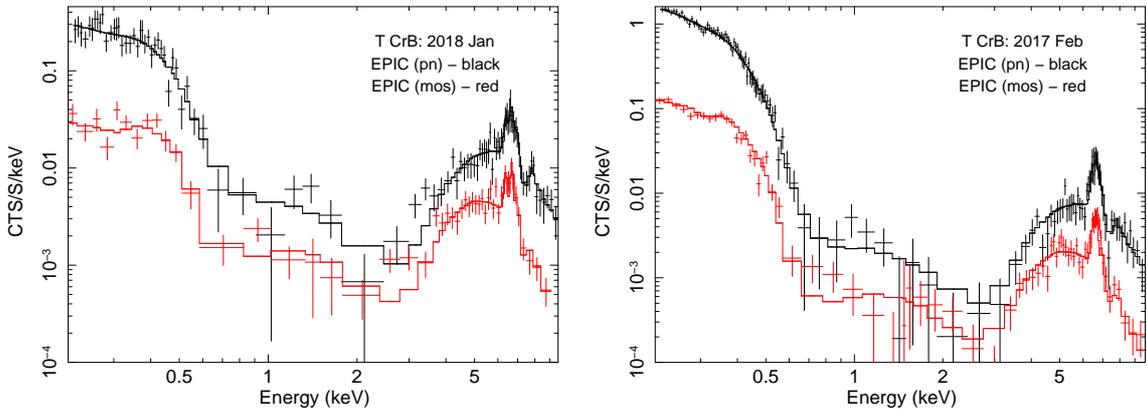

\begin{center}
\includegraphics[width=2.1in, height=3.0in, angle=-90]{fig3a.eps}
\includegraphics[width=2.1in, height=3.0in, angle=-90]{fig3b.eps}
\end{center}
\caption{The background-subtracted spectra of \TCRB from the \XMM
observations in 2018 (left panel) and 2017 (right panel) overlaid
with a two-component thermal  model (Model D; Table~\ref{tab:fits}).
Spectra are re-binned to have a minimum of 30 counts per bin.
}
\label{fig:spec_xmm}
\end{figure*}

\begin{table*}
\caption{Global Spectral Model Results
\label{tab:fits}}
\begin{tabular}{lllllllll}
\hline
\multicolumn{1}{c}{Parameter} & 
\multicolumn{4}{c}{BB + vmcflow }  & 
\multicolumn{4}{c}{BB + vapec }  \\
\multicolumn{1}{c}{ } &
\multicolumn{2}{c}{Model A}  & \multicolumn{2}{c}{Model B} &
\multicolumn{2}{c}{Model C}  & \multicolumn{2}{c}{Model D} \\
\multicolumn{1}{c}{ } &
\multicolumn{1}{c}{2017}  & \multicolumn{1}{c}{2018} &
\multicolumn{1}{c}{2017}  & \multicolumn{1}{c}{2018} &
\multicolumn{1}{c}{2017}  & \multicolumn{1}{c}{2018} &
\multicolumn{1}{c}{2017}  & \multicolumn{1}{c}{2018} \\
\hline
$\chi^2$/dof  & 
          \multicolumn{2}{c}{309/335}  & \multicolumn{2}{c}{326/338}  & 
          \multicolumn{2}{c}{308/335}  & \multicolumn{2}{c}{324/338}  \\
N$_{H,1}$ (10$^{20}$ cm$^{-2}$)  &
          \multicolumn{2}{c}{5.19$^{+0.59}_{-0.45}$} & 
          \multicolumn{2}{c}{5.27$^{+0.46}_{-0.32}$} & 
          \multicolumn{2}{c}{5.17$^{+0.34}_{-0.26}$} & 
          \multicolumn{2}{c}{5.18$^{+0.40}_{-0.39}$} \\
CF  &
          0.997$^{+0.001}_{-0.001}$ & 0.995$^{+0.001}_{-0.001}$ &
          \multicolumn{2}{c}{0.996$^{+0.001}_{-0.001}$} & 
          0.995$^{+0.001}_{-0.001}$ & 0.993$^{+0.001}_{-0.001}$ &
          \multicolumn{2}{c}{0.994$^{+0.001}_{-0.001}$} \\
N$_{H,2}$ (10$^{22}$ cm$^{-2}$)  &
          43.3$^{+2.79}_{-2.39}$ & 33.1$^{+2.13}_{-1.84}$ & 
          \multicolumn{2}{c}{38.2$^{+1.41}_{-1.63}$} & 
          42.8$^{+2.44}_{-2.30}$ & 33.1$^{+1.89}_{-1.78}$ & 
          \multicolumn{2}{c}{38.1$^{+1.50}_{-1.43}$} \\ 
kT$_{BB}$ (keV) & 
          0.035$^{+0.001}_{-0.001}$ & 0.045$^{+0.002}_{-0.002}$ & 
          0.035$^{+0.001}_{-0.001}$ & 0.045$^{+0.001}_{-0.001}$ & 
          0.035$^{+0.001}_{-0.001}$ & 0.045$^{+0.002}_{-0.002}$ & 
          0.035$^{+0.001}_{-0.001}$ & 0.045$^{+0.002}_{-0.002}$ \\ 
R$_{BB}$ (km) & 
          2285$^{+464}_{-365}$ & 281$^{+65}_{-46}$ & 
          2063$^{+268}_{-440}$ & 322$^{+68}_{-25}$ & 
          1794$^{+434}_{-230}$ & 233$^{+36}_{-37}$ & 
          1631$^{+269}_{-226}$ & 255$^{+40}_{-34}$ \\ 
Fe      & 
          \multicolumn{2}{c}{0.97$^{+0.05}_{-0.07}$} &
          \multicolumn{2}{c}{0.92$^{+0.12}_{-0.13}$} &
          \multicolumn{2}{c}{0.84$^{+0.11}_{-0.06}$} &
          \multicolumn{2}{c}{0.80$^{+0.09}_{-0.11}$} \\
Ni      & 
          \multicolumn{2}{c}{4.62$^{+1.62}_{-1.61}$} &
          \multicolumn{2}{c}{4.50$^{+1.69}_{-1.67}$} &
          \multicolumn{2}{c}{3.85$^{+1.44}_{-1.47}$} &
          \multicolumn{2}{c}{3.82$^{+1.46}_{-1.45}$} \\
kT$_{max}$ (keV) & 
          11.28$^{+1.56}_{-0.58}$ & 18.52$^{+3.53}_{-3.22}$ & 
          \multicolumn{2}{c}{14.67$^{+2.24}_{-1.27}$} & 
                &   &   &   \\
$\dot{M}$ ($10^{-10}$\dotM) &
          1.71$^{+0.32}_{-0.35}$ & 1.46$^{+0.33}_{-0.26}$ &
          1.09$^{+0.21}_{-0.20}$ & 2.15$^{+0.45}_{-0.38}$ &
                &   &   &   \\
kT (keV) &
                &   &   &   &
          6.63$^{+0.39}_{-0.44}$ & 8.26$^{+0.53}_{-0.44}$ &
          \multicolumn{2}{c}{7.53$^{+0.48}_{-0.24}$} \\
EM ($10^{55}$~cm$^{-3}$) &  
                                 & & & &
                                1.41$^{+0.18}_{-0.14}$ &
                                1.98$^{+0.17}_{-0.16}$ &
                                1.18$^{+0.09}_{-0.08}$ &
                                2.35$^{+0.19}_{-0.14}$ \\
E$_{l}$ (keV) &
       6.45$^{+0.02}_{-0.01}$ & 6.38$^{+0.03}_{-0.01}$ &
       6.48$^{+0.01}_{-0.02}$ & 6.40$^{+0.02}_{-0.02}$ &
       6.46$^{+0.01}_{-0.02}$ & 6.39$^{+0.02}_{-0.01}$ &
       6.48$^{+0.02}_{-0.01}$ & 6.38$^{+0.02}_{-0.02}$ \\
F$_{l}$ ($10^{-6}$ cts cm$^{-2}$ s$^{-1}$) &
       3.50$^{+0.55}_{-0.55}$ & 6.64$^{+0.97}_{-0.97}$ &
       4.00$^{+0.66}_{-0.51}$ & 6.29$^{+1.05}_{-0.95}$ &
       3.52$^{+0.66}_{-0.47}$ & 6.64$^{+0.98}_{-0.94}$ &
       4.08$^{+0.57}_{-0.57}$ & 6.30$^{+0.96}_{-0.96}$ \\
F$_{X}$ (0.2 - 10 keV)  &
           \,\,1.29  & \,\,1.35  &
           \,\,1.28  & \,\,1.35  &
           \,\,1.30  & \,\,1.35  &
           \,\,1.29  & \,\,1.36  \\
                                              &
            (2067) &  (187) &
            (1641) &  (238) &
            (1288) &  (131) &
            (1059) &  (156) \\
F$_{X}$ (0.2 - 0.6 keV)  &
           \,\,0.66  & \,\,0.14  &
           \,\,0.66  & \,\,0.14  &
           \,\,0.66  & \,\,0.14  &
           \,\,0.66  & \,\,0.14  \\
                                              &
            (2062) &  (181) &
            (1637) &  (230) &
            (1284) &  (126) &
            (1056) &  (150) \\
$\log L_{BB}$ (erg s$^{-1}$) &
           35.99 & 34.63 & 35.90 & 34.73 &
           35.78 & 34.46 & 35.70 & 34.54 \\
$\log L_{BB}^{nc}$ (erg s$^{-1}$) &
           33.45 & 32.29 & 33.46 & 32.30 &
           33.44 & 32.28 & 33.44 & 32.29 \\
$\log L_{H}$ \,\,\,\,(erg s$^{-1}$) &
           32.64 & 32.81 & 32.57 & 32.87 &
           32.54 & 32.72 & 32.36 & 32.66 \\
\hline

\end{tabular}

{\it Note}.
Results from  fits to the EPIC spectra of \TCRB obtained in 2017
February and 2018 January.  The \xspec models are: 
$wabs((partcov*wabs)(bbodyrad+vmcflow) + gaussian)$ (Model A); 
the same as Model A but the parameters CF, N$_{H,2}$ and kT$_{max}$ 
have the same values for the 2017 and 2018 spectra (Model B);
$wabs((partcov*wabs)(bbodyrad+vapec) + gaussian)$ (Model C); 
the same as Model C but the parameters CF, N$_{H,2}$ and kT
have the same values for the 2017 and 2018 spectra (Model D);
Tabulated quantities are the neutral hydrogen absorption column
density (N$_{H,1}$; representative of the interstellar absorption), 
the covering factor of the partial-covering absorption (CF),
hydrogen absorption column density (N$_{H,2}$; associated with
material in the \TCRB system), black-body temperature (kT$_{BB}$),
radius of the black body (R$_{BB}$),
iron and nickel abundances of the plasma component (Fe, Ni),
the maximum plasma temperature in the cooling flow (kT$_{max}$),
normalization parameter (mass-loss rate) for the cooling flow
($\dot{M}$), plasma temperature of the hard component (kT) and its
emission measure ($\mbox{EM} = \int n_e n_{H} dV $),
central energy (E$_{l}$) and its flux (F$_{l}$) of the line component,
the observed X-ray fluxes 
(F$_X$) 
in the (0.2 - 10 keV) and (0.2 - 0.6
keV) ranges followed in parentheses by their unabsorbed values
(units are $10^{-12}$ erg cm$^{-2}$ s$^{-1}$), 
the luminosity of the black-body component ($L_{BB}$) and its value if
this component is subject only to the interstellar absorption 
($L_{BB}^{nc}$, see text for details), the luminosity of the hard
component $L_{H}$ in the (0.2 - 80 keV) range.
The derived abundances are with respect to the solar abundances
\citep{ag_89}.
The values for the emission measure and the X-ray luminosity are for a
reference distance of d~$=806$~ pc.
Errors are the $1\sigma$ values from the fits.

\end{table*}

\subsection{Global Spectral Models}
\label{sec:xray_spec}

As discussed in Section~\ref{sec:xray_overview}, we need at least two
components in the global spectral models to derive some physical 
properties of the X-ray emission from \TCRBE. However, if we adopt
just two spectral components that are supposed to correspondingly  
match the strong soft emission (0.2 - 0.6 keV) and the highly absorbed
hard emission (E $> 2$~keV), such a model cannot match the weak X-ray 
emission at 0.6 - 2 keV (see Fig.~\ref{fig:pn_spectra}, left panel). 
The (0.6 - 2 keV) emission can be fitted successfully if we introduce 
an additional spectral component. Unfortunately, due to the limited 
photon statistics of this emission the model parameters of the third 
component cannot be constrained. 

We recall that the X-ray emission in the (0.6 - 2 keV) range
correlates with the highly absorbed emission at energies E $> 2$~keV
(see above), and indeed our attempts to find a common spectral model 
for it and the strong soft emission (0.2 - 0.6 keV) ended with no 
success. So, one way of resolving this issue is to adopt the approach 
proposed by \citet{luna_18}, namely, to consider a partial-covering 
factor for the highly absorbed hard component. 

First, we considered a spectral model with partial-covering absorption
as in \citet{luna_18} that consists of the following components: black 
body and cooling flow emission as basic ingredients, and a gaussian 
component, representing  the Fe K-shell fluorescent line in the
spectrum of \TCRBE.
Second, we considered a similar model whose second spectral
component is that of optically thin plasma emission.
Such a choice is justified by the fact that the physical picture of
cooling flow (typical for clusters of galaxies) is not adequate to 
that in an accreting binary system\footnote{
We recall that cooling flows (CF) in clusters of galaxies are assumed 
to be spherically symmetric and such is the CF model in 
\xspecE. But, the gas flow  in binary systems is {\it not}: valid both 
for accretion from stellar wind and mass transfer 
due to Roche lobe overflow.
This is also the case with the accretion disk boundary layer which is 
{\it not} a spherically-symmetric object, but its emission is
supposed to be represented by the CF model
(see Section~\ref{sec:discussion}).
}
no matter whether such a model 
(with a cooling flow component) may provide acceptable fits to the 
observed spectra.

We fitted all the observed spectra (EPIC 2018 and 2017) simultaneously 
as the pn and MOS spectra of each individual data set shared identical
model parameters.
The chemical abundances were solar \citep{ag_89} with no evolution
between 2017 and 2018.
To improve the quality of the fits, the iron and nickel (Fe and Ni) 
abundances  were allowed to vary.
Table~\ref{tab:fits} and Fig.~\ref{fig:spec_xmm} present the
corresponding results from our fits to the \XMM spectra of \TCRBE.

Since we did not find very big differences in some of the model
parameters between 2017 and 2018 (see Model A and C in
Table~\ref{tab:fits}), we considered models in which these parameters
had the same values for the 2017 and 2018 spectra. Such were the
partial-covering absorption and the temperature of the hard component
(or the maximum temperature in the cooling flow). The quality of these
fits (see Model B and D in Table~\ref{tab:fits}) is very good and
justifies such an assumption. On the other hand, our attempts to find
acceptable fits with models that assumed the same black-body 
temperature for the 2017 and 2018 spectra were not successful. 

We also explored models in which the black-body emission was subject
{\it only} to the interstellar absorption, thus, the partial-covering
absorption was related to the hard component. These models gave the
same quality of the fits and the same model parameters
as those whose results are given in
Table~\ref{tab:fits}. The only difference was in the value of the
black-body luminosity, which is given therein: see parameter
$L_{BB}^{nc}$. 

As seen from the spectral fit results (Table~\ref{tab:fits}), 
nickel seems to be overabundant compared to iron. However, caution is
advised before proposing any physical reason as explanation. It is so
since this result comes from analysis of low-resolution CCD spectra,
it is based on {\it only} one spectral feature (Ni XXVII at $\sim
7.8$~ keV) and its formal statistical error is not low (it is just a 
$2\sigma$ result).

Finally, we note that the values of the hydrogen column density of the 
interstellar absorption derived from the fits (see parameter N$_{H,1}$ 
in Table~\ref{tab:fits}) is consistent (within 10\%) with the Galactic
H I column density towards
\TCRBE\footnote{\url{https://heasarc.gsfc.nasa.gov/cgi-bin/Tools/w3nh/w3nh.pl}
; based on \citet{map1} and \citet{map2}} and
these X-ray derived values are equivalent to a 
range of optical extinction A$_{\mathrm{V}} = 0.23 - 0.32$ mag. 
The range corresponds to the conversion that is used:
N$_{\mathrm{H}} = 2.22\times10^{21}$A$_{\mathrm{V}}$~cm$^{-2}$ \citep{go_75}
and 
N$_{\mathrm{H}} = (1.6-1.7)\times10^{21}$A$_{\mathrm{V}}$~cm$^{-2}$
(\citealt{vuong_03}, \citealt{getman_05}).

\begin{figure}
\begin{center}
\includegraphics[width=3.0in, height=2.145in]{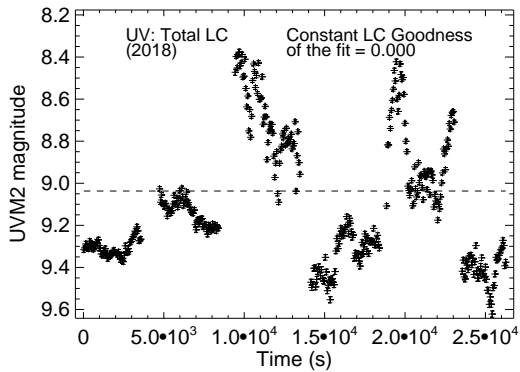}
\end{center}
\caption{The UV light curves of \TCRB from the \XMM OM (optical
monitor) observations in 2018 January. The LCs were binned at 60 s.
The dashed line denotes the constant (mean) flux.
}
\label{fig:lc_uv}
\end{figure}

\begin{figure*}
\begin{center}
\includegraphics[width=3.0in, height=2.145in]{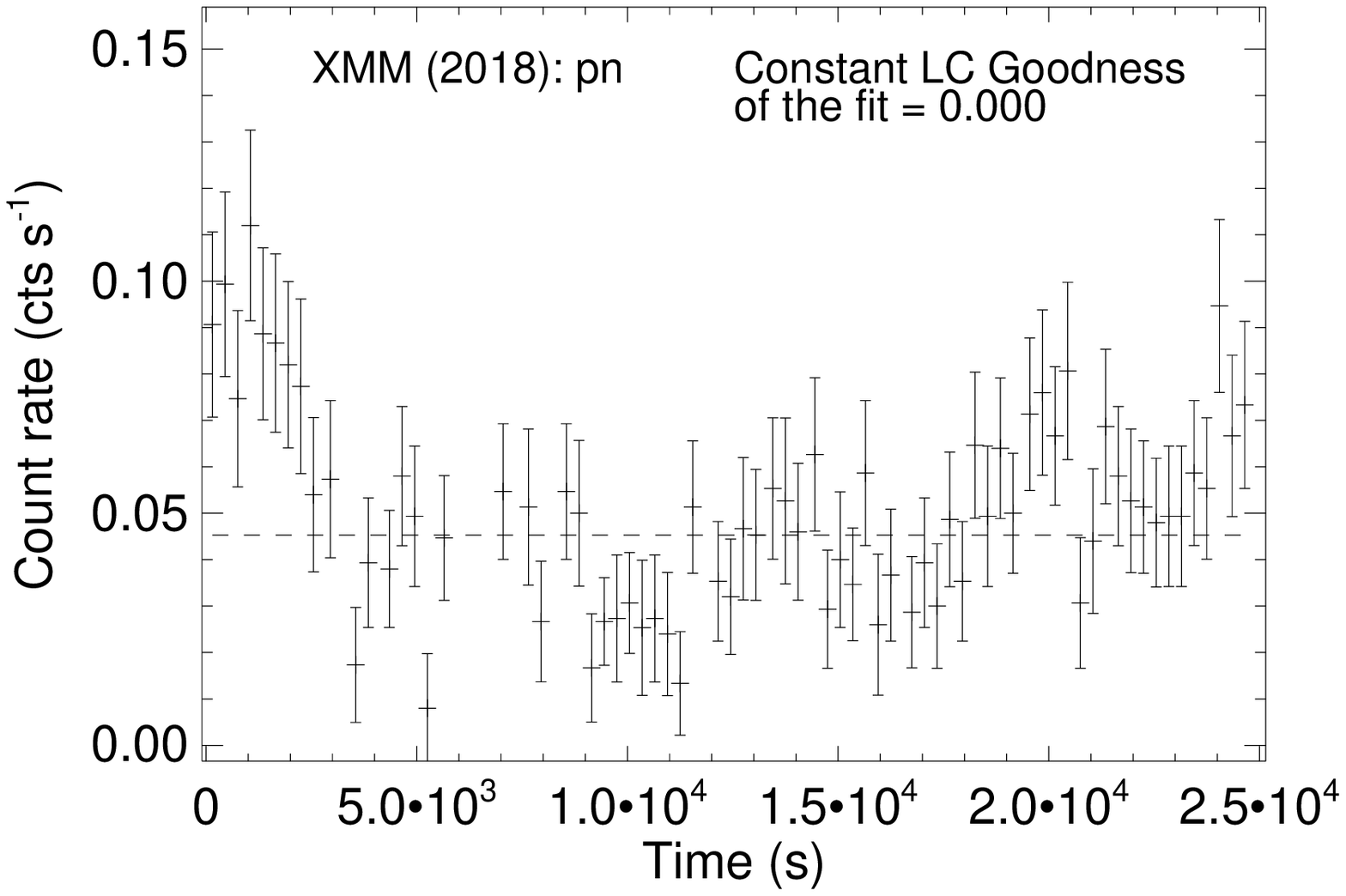}
\includegraphics[width=3.0in, height=2.145in]{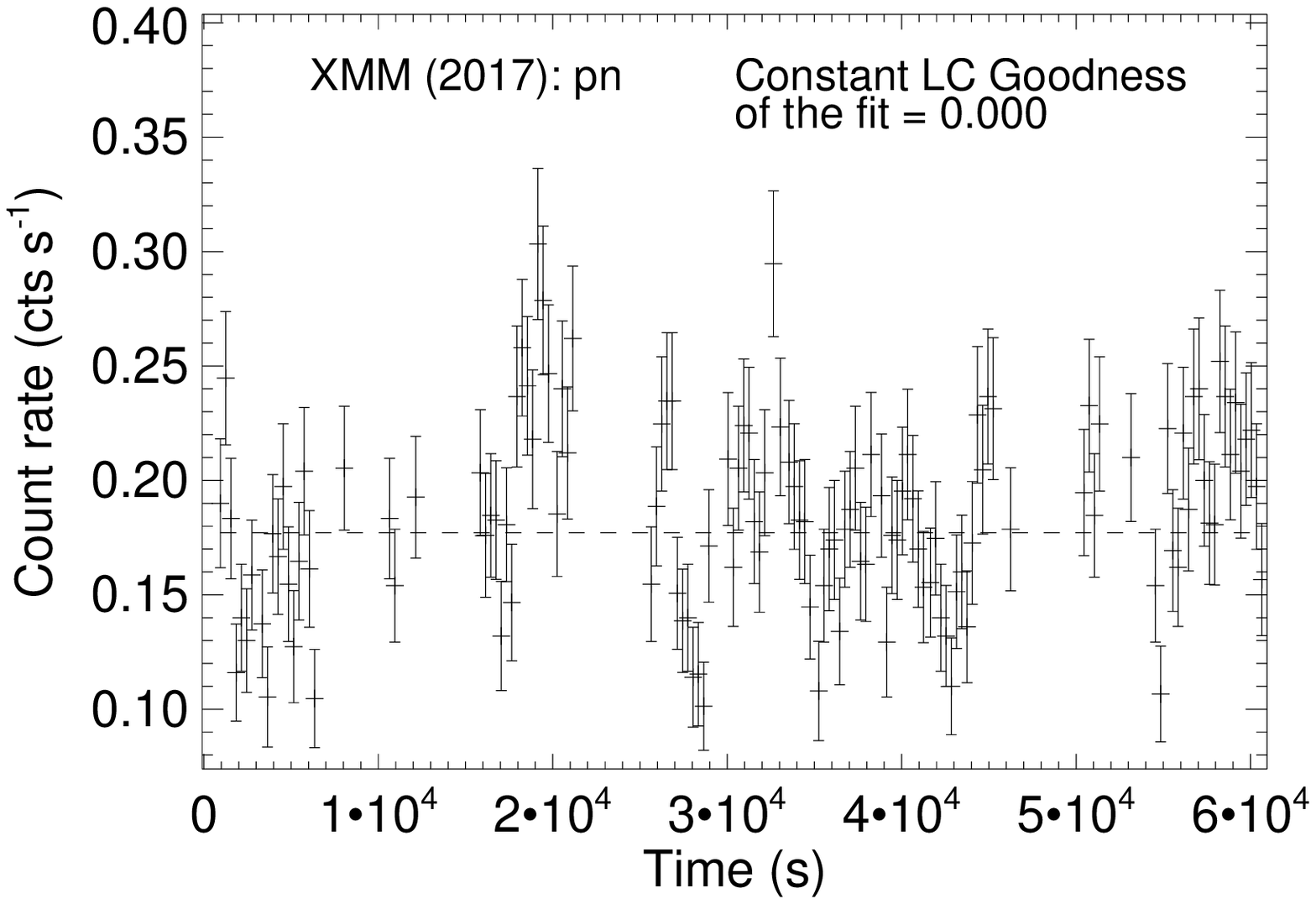}
\includegraphics[width=3.0in, height=2.145in]{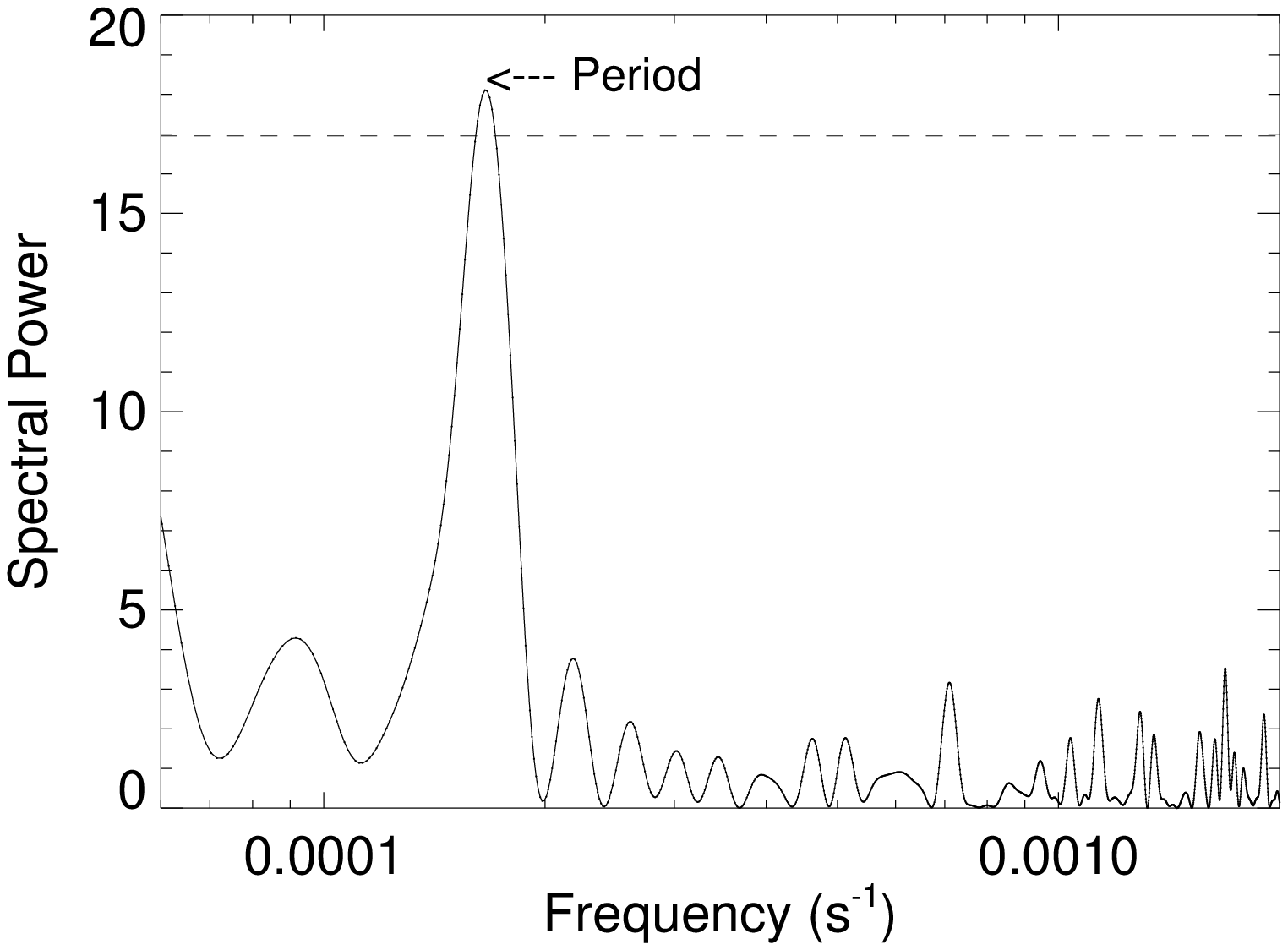}
\includegraphics[width=3.0in, height=2.145in]{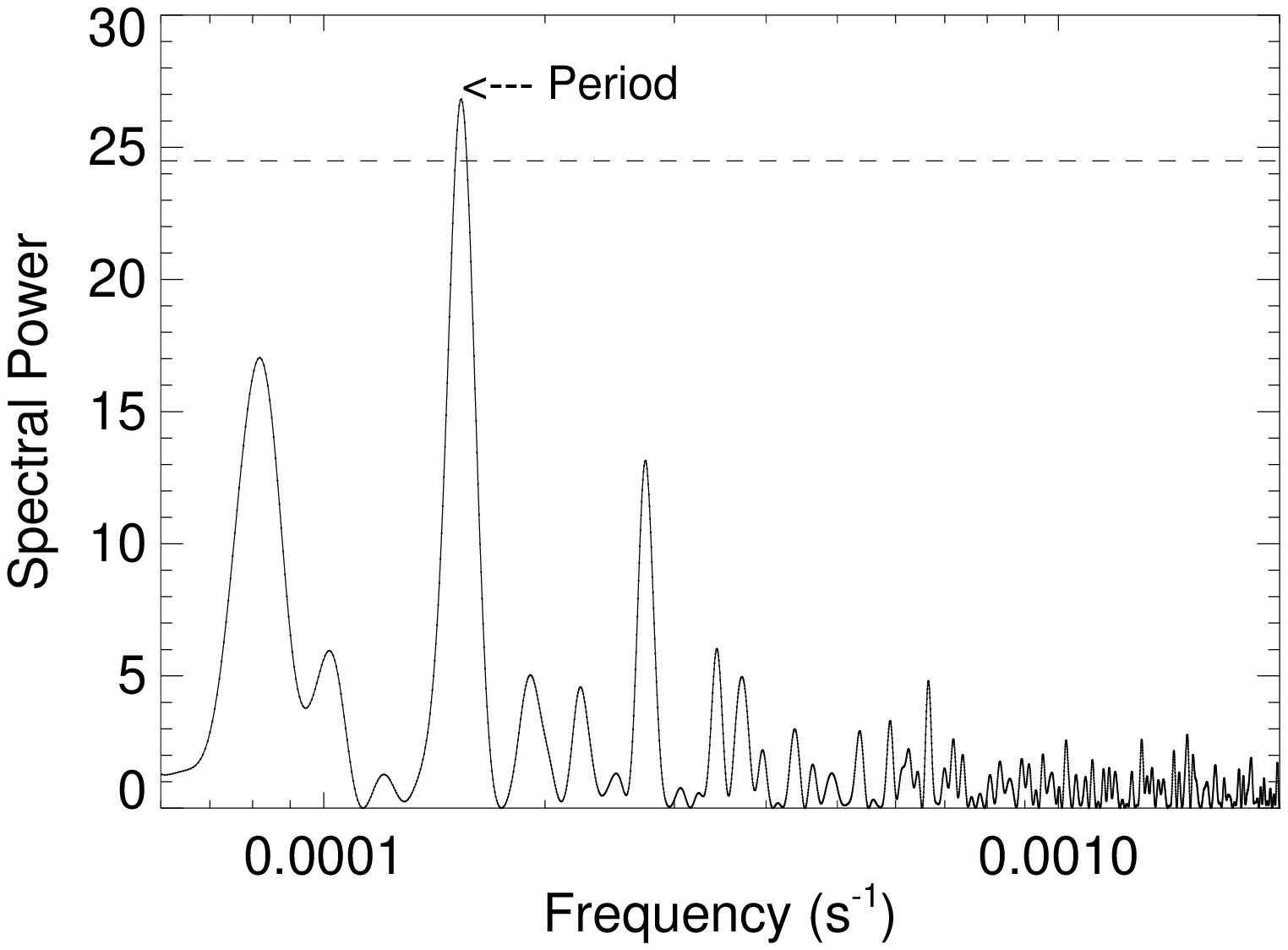}
\includegraphics[width=3.0in, height=2.145in]{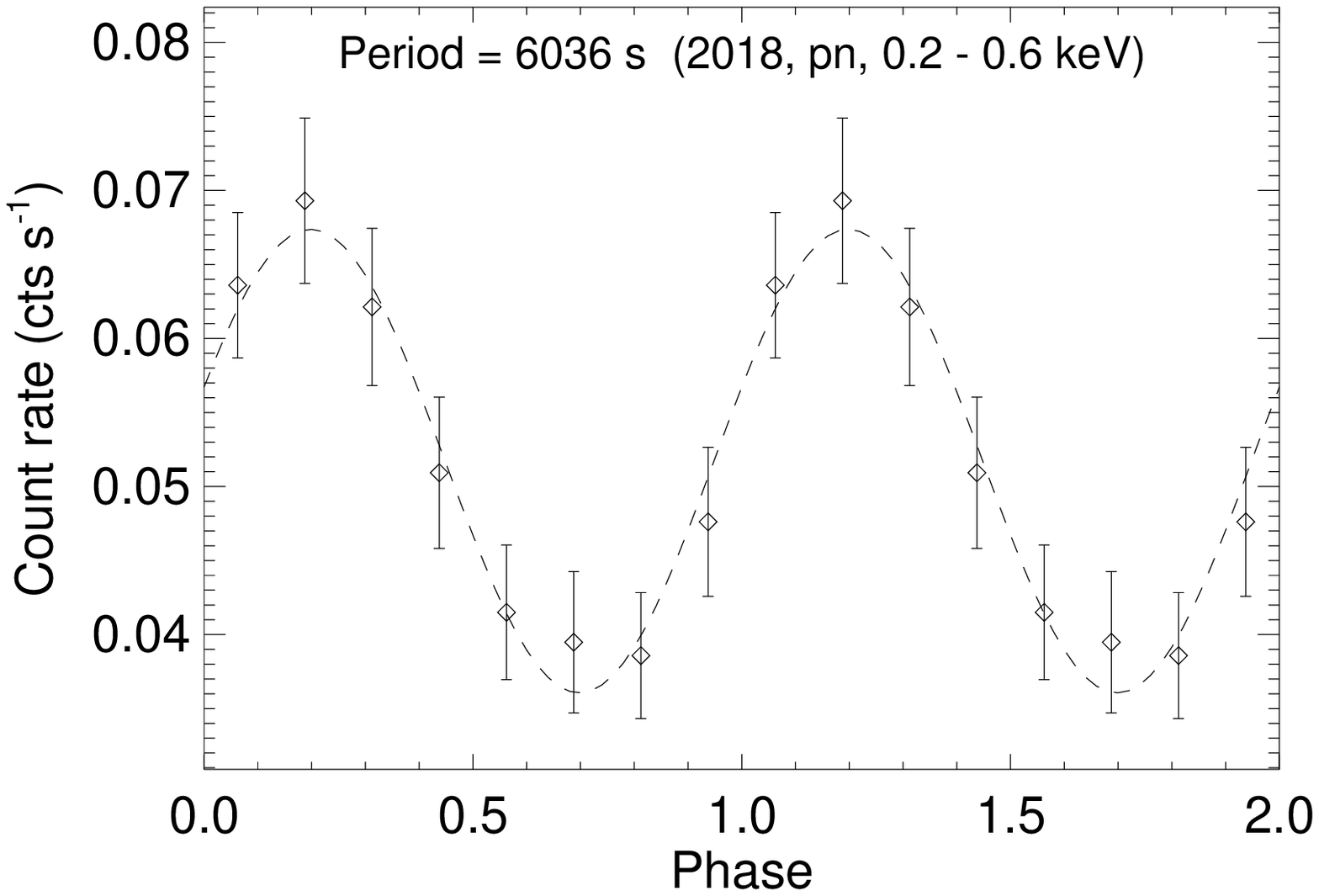}
\includegraphics[width=3.0in, height=2.145in]{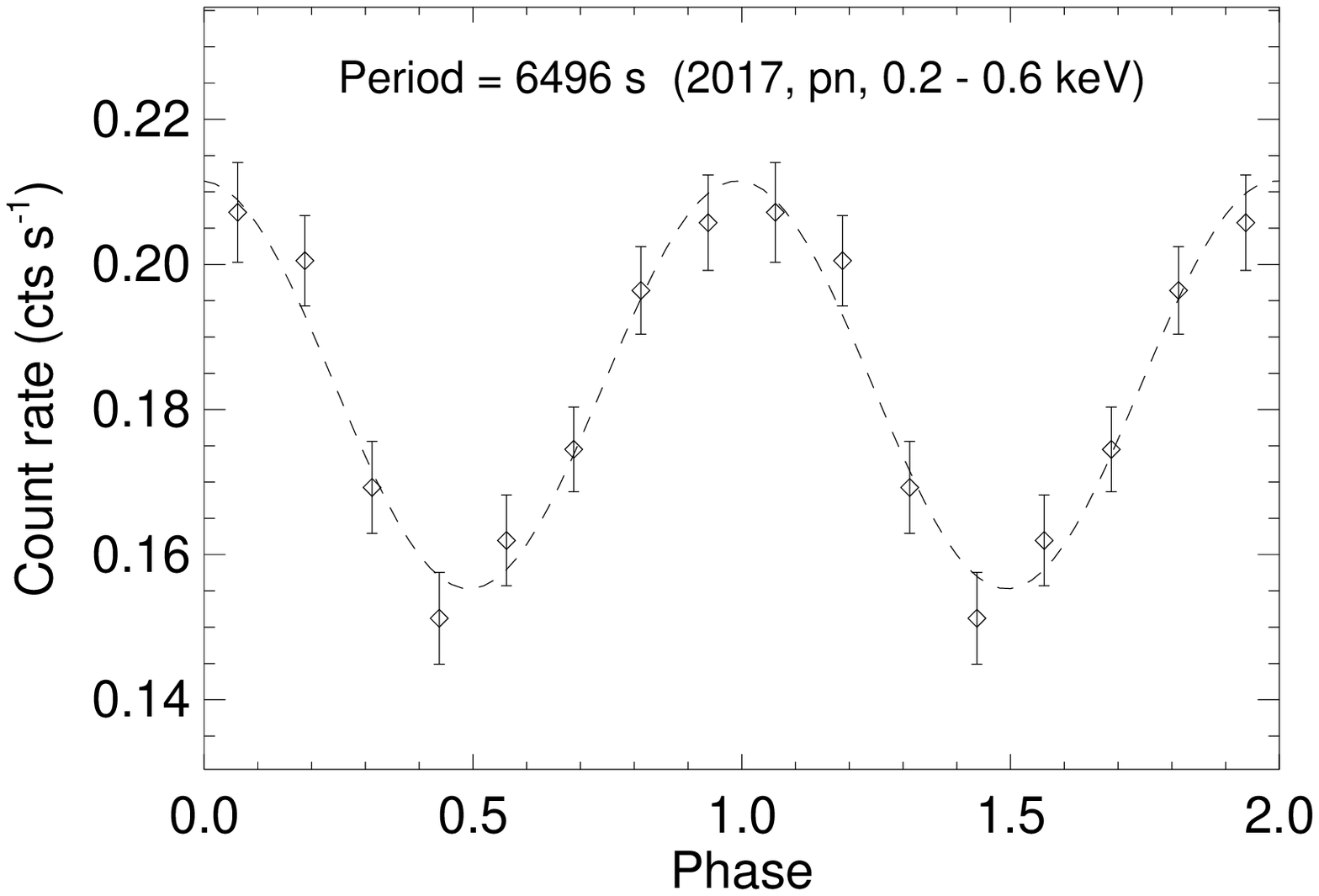}
\includegraphics[width=3.0in, height=2.145in]{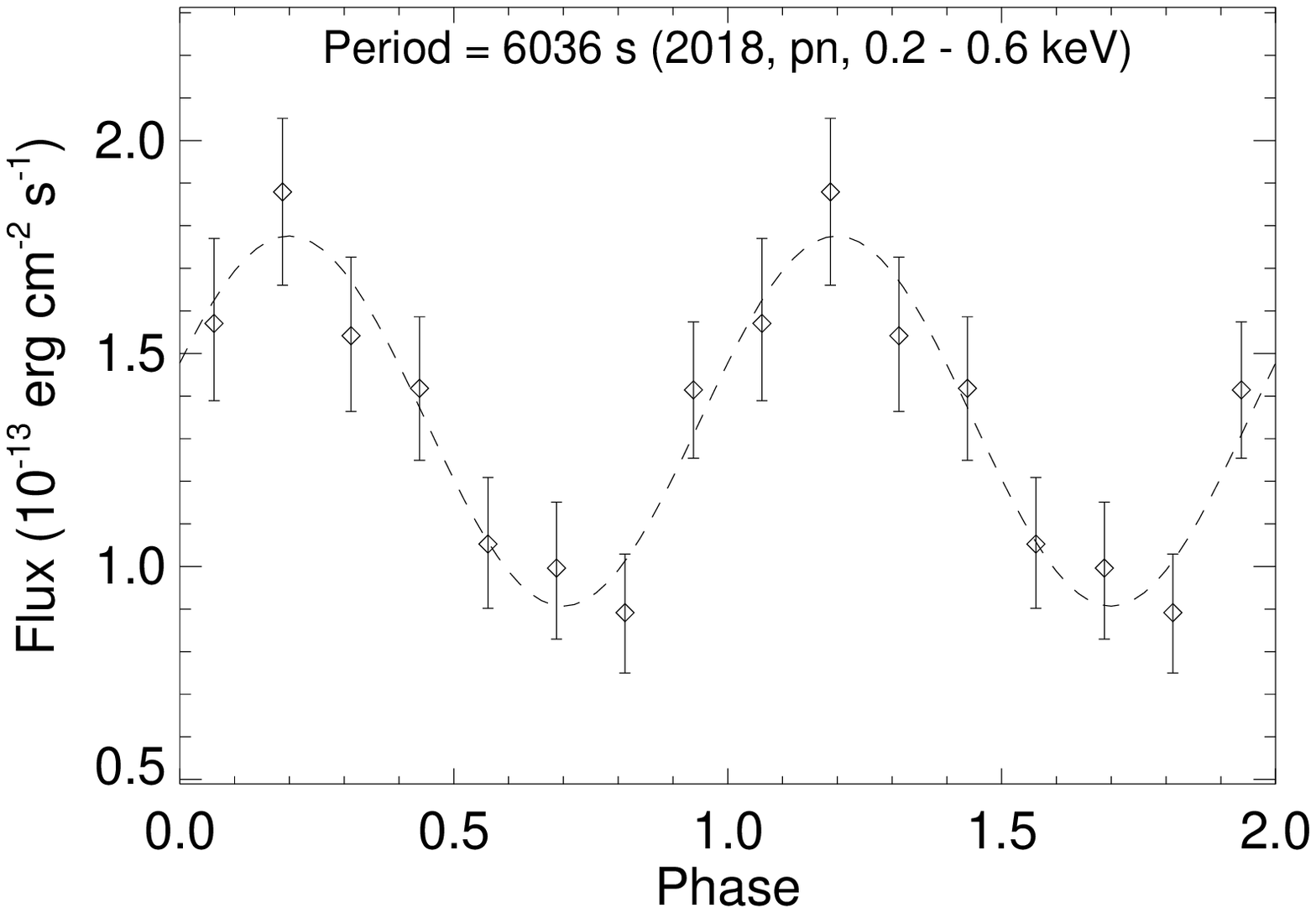}
\includegraphics[width=3.0in, height=2.145in]{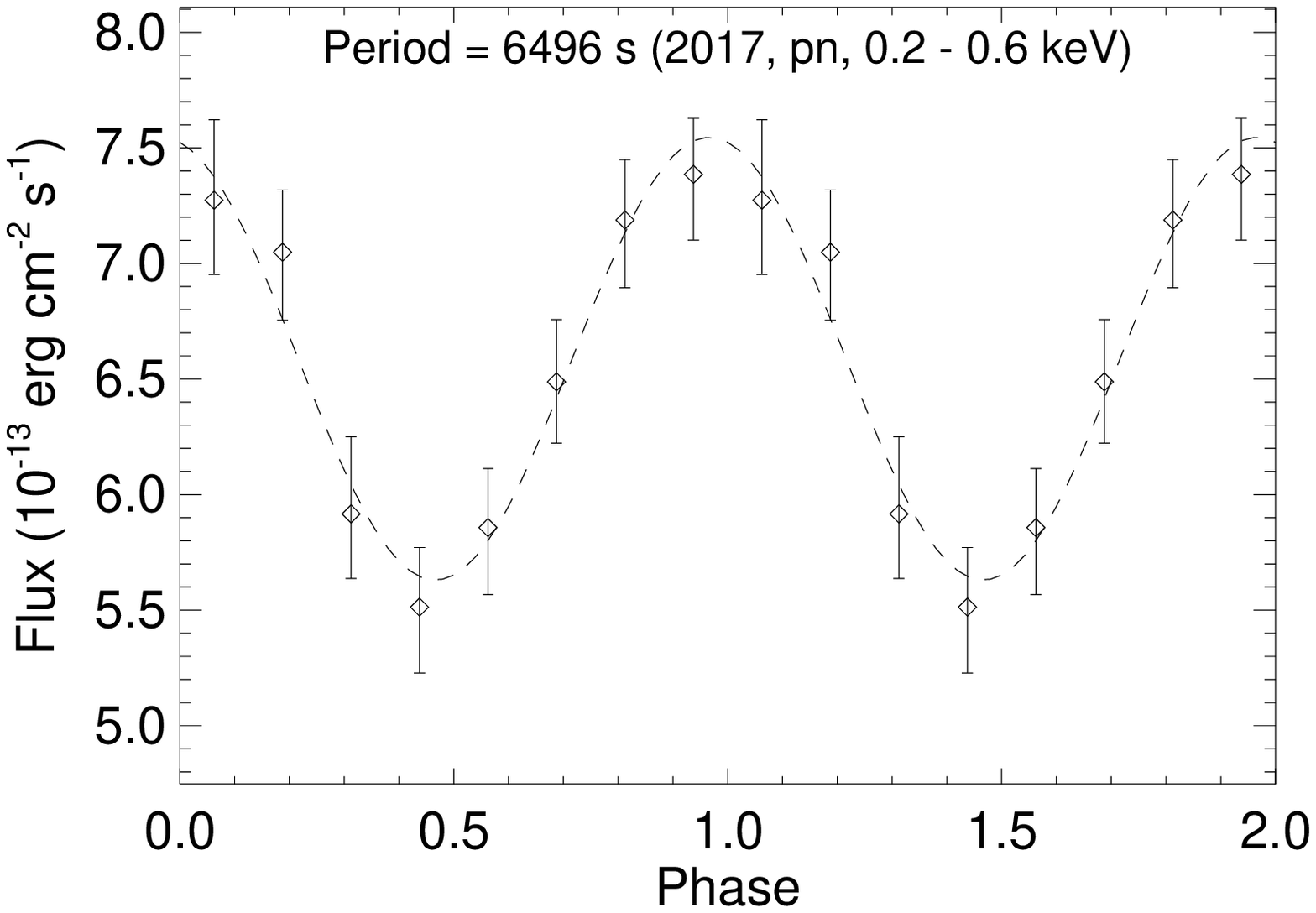}
\end{center}
\caption{\XMM light curves of T CrB. 
The LCs  with a 100-s time bin are shown in the first row
(slightly re-binned for presentation).
The Lomb-Scargle power spectra are shown in
the second row:
the dashed line marks to the false-alarm probability 
level of $10^{-5}$ (2018) and $10^{-8}$ (2017). 
The X-ray LCs folded with the derived period are
given in the third row: each LC points presents the average count
rate within a phase bin with a size of 0.125. The forth row presents
the X-ray LCs in flux units (see text for details).
The fit with a simple sinusoidal curve (with the noted period) is
given by the dashed line.
}
\label{fig:period}
\end{figure*}

\subsection{Variability}
\label{sec:xray_lc}

Previous studies of \TCRB  showed that it possess strong stochastic
variability in the X-ray and UV domains (e.g., \citealt{luna_08}; 
\citealt{ilkiewicz_16}; \citealt{luna_18}). 
The \XMM observations of \TCRB allow to check whether such a
variability is present during the current active phase.
To do so, we fitted each LC with a constant, adopting
$\chi^2$ fitting.

Figure~\ref{fig:lc_uv} shows the total UV light curve for the 2018 
data set in the UVM2 filter of the  \XMM OM (optical monitor). We note
that the 2018 observation had six individual exposures 
(of $\sim 3000$~s each).  We explored different binning
of the LCs but no matter the time-bin size, stochastic variability
({\it flickering}) was always present.
This result does not depend on the format of the light
curve: it is valid for the LCs in flux units (counts s$^{-1}$) and in
magnitudes.

In X-rays, we made use of the background-subtracted light curves from 
all the EPIC (pn, MOS1 and MOS2) detectors in the soft (0.2 - 0.6
keV), hard (2 - 10 keV) and total (0.2 - 10 keV) energy bands,
adopting various binning with time bins between 100 and 1000 s.
In all the cases under consideration, {\it flickering} was present 
both in the 2018 and 2017 data sets. The higher sensitivity of the pn 
detector results in LCs with much better quality compared to those 
from the MOS detectors, even if the latter were combined (MOS1+MOS2).
This allowed us to search for periodic signal in the X-ray light 
curves of \TCRBE.

For this purpose, we 
applied the Lomb-Scargle method (\citealt{lomb_76}; 
\citealt{scargle_82}; see \S~13.8 in \citealt{press_92})
to the pn LCs of \TCRBE, since the \XMM LCs are typically unevenly 
sampled data due to time intervals with high X-ray background.
We analysed the soft, hard and total X-ray LCs. The Lomb-Scargle
power spectrum indicated that a periodic signal was present in the
soft (0.2 - 0.6 keV) emission of \TCRB as the false-alarm probability
was very low ($< 10^{-5}$), that is, the signal is highly significant:
the smaller the value the more significant the signal (see \S~13.8 in
\citealt{press_92}).

To check this result,
we constructed a phase-averaged light curve by folding the original LC 
with the found period as the count rate was averaged within the size 
of the phase bin to minimize the contribution from the flickering.
Such phase-averaged LCs could be very well matched by a simple
sinusoidal function.

As an additional check, we performed time-resolved spectroscopy of the
X-ray emission from \TCRBE. Namely, we re-extracted the pn spectra for
each phase bin by combining all the X-ray emission with exposure time
that exactly corresponds to that particular phase bin. 
We note that thanks to the good photon statistics of the EPIC-pn data
of \TCRB the individual phase-bin spectra have enough source counts
($\geq 270$ for the 2018 and $\geq 820$ for the 2017 spectra,
respectively) which allows for their analysis in some detail. So, we 
fitted the phase-bin spectra with
the same global models as discussed in Section~\ref{sec:xray_spec}.
In these fits, we varied only the normalization parameters of the 
spectral components, while the other parameters were kept fixed to 
their values given in Table~\ref{tab:fits}. We found that the
flux variations of the soft X-ray emission from \TCRB well correspond
to the periodic signal detected by applying the Lomb-Scargle method to
the X-ray LCs from the 2018 and 2017 \XMM observations.

Some results from our analysis of the soft X-ray emission of \TCRB
aimed at searching for periodic signal are given in 
Fig.~\ref{fig:period}. We think it is conclusive that periodic signal 
(with a period of 6000 - 6500 s) was very likely present in the 
{\it soft} X-ray emission from the recurrent symbiotic nova \TCRB in 
its active phase that started in 2014-2015.

We note that presence of flickering may result in suggesting some
false signals from periodogram analysis. The result may depend on
relative strength of the flicker noise and the real periodic signal.
Some technical efforts might be needed to estimate the statistical
significance of periodic signals, especially, those with short periods
less than 1000 s or so (e.g., see \citealt{neustroev_05} and 
references therein).

However, we underline that our conclusion of the peioric 
(6000 - 6500 s) signal in the soft X-ray emission of \TCRB is not just 
statistical and it is based 
on the accumulative evidence from the discussed above findings. 
Namely, (a) the Lomb-Sargle method indicates a periodic
signal; (b) the light curves folded with the suggested period do
confirm its presence; (c) the suggested period is also confirmed by 
time-resolved spectroscopy.
However, the most important fact is the {\it repeatability} of the 
periodic signal: it is 
present in two different observational data sets, taken almost a year
apart from one another, and the corresponding  values of the period
differ by less than 10\%. 

Finally, we also checked the correlation between the soft (0.2 - 0.6
keV) and hard (2 - 10 keV) emission in the 2018 and 2017 \XMM LCs of
\TCRBE. The  value of the linear Pearson correlation coefficient 
between the soft and hard LCs is small both for the 2018 and 2017 
observations ($r \approx -0.07; -0.1$), thus, indicating very weak 
(or even no) correlation.
Such a different behaviour of the X-ray LCs likely means that the 
soft and hard X-ray emissions originate from different regions in 
\TCRB and might be subject to different formation mechanisms.

\begin{figure*}
\begin{center}
\includegraphics[width=3.0in, height=2.145in]{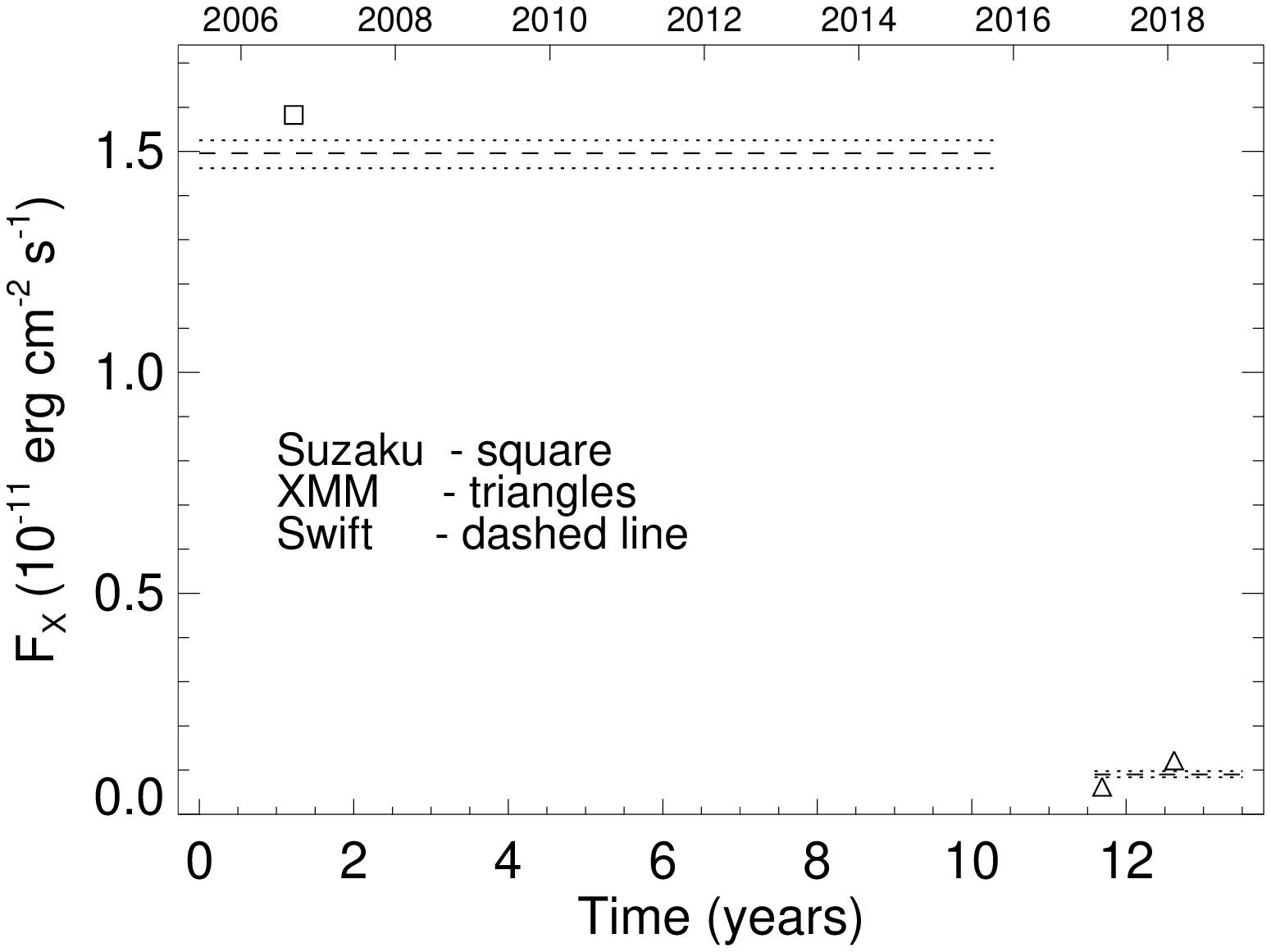}
\includegraphics[width=3.0in, height=2.145in]{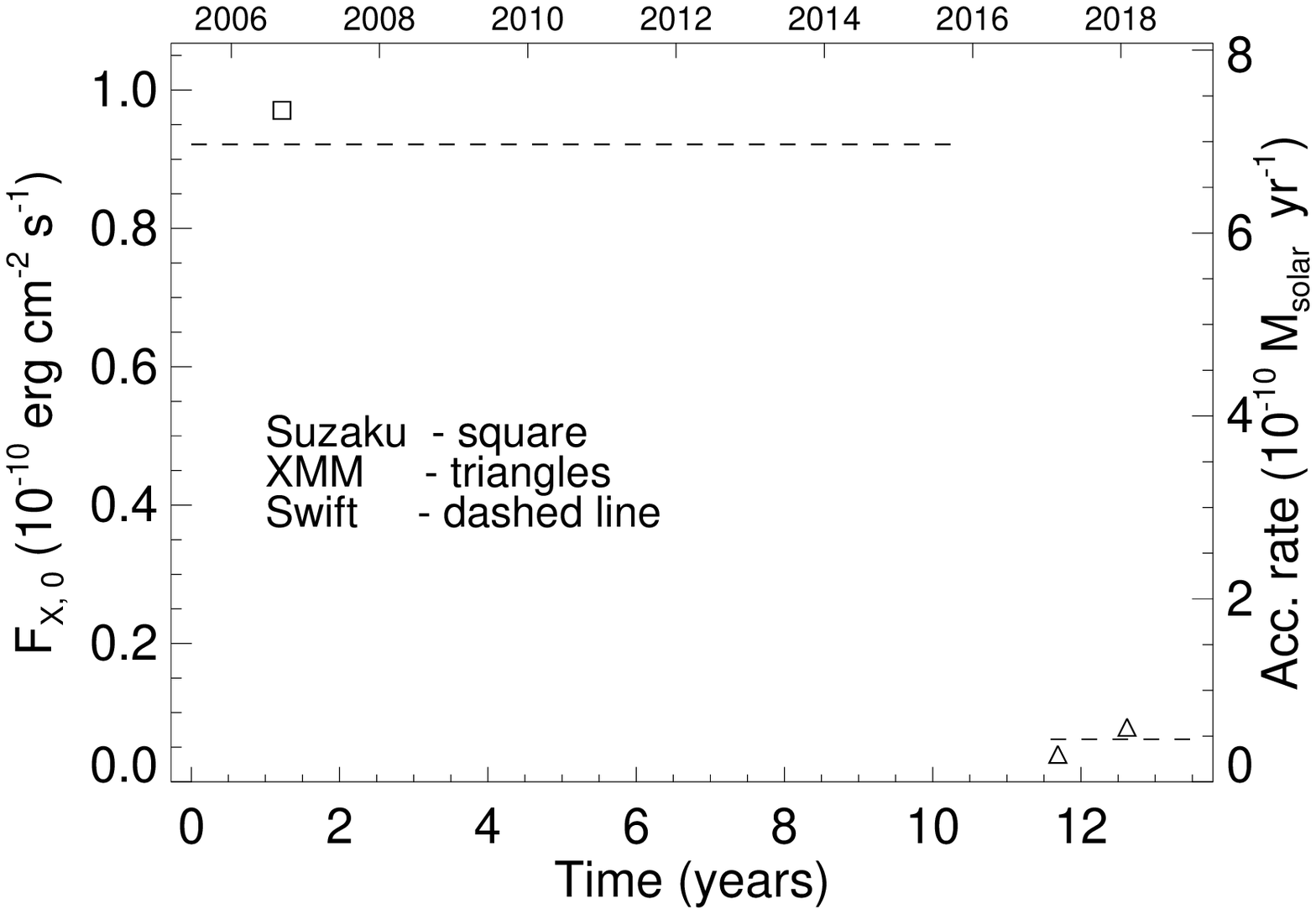}
\end{center}
\caption{The long-term light curve of \TCRB from observations
with some X-ray observatories.
{\it Left} panel: the observed flux in the 1.5 - 10 keV energy range.
The dotted lines mark the $1\sigma$-confidence range on the \Swift
flux, while that for the \Suzaku and \XMM fluxes is within the size of
the corresponding symbols denoting the flux values.
{\it Right} panel: the intrinsic (unabsorbed) flux in the 0.2 - 80 keV
energy range of the hard component in the X-ray spectrum
(the corresponding accretion rate is shown on the right-hand
side y-axis).
}
\label{fig:lc_suz_swi}
\end{figure*}

\section{Discussion}
\label{sec:discussion}

The \XMM observations of \TCRB in 2018 January and 2017 February 
provided X-ray data with good quality. One of the most important 
results from our analysis of these data is the evolution of the X-ray 
emission from this symbiotic recurrent nova during its active phase.

We recall (see above) that the X-ray emission from \TCRB has two basic
(prominent) components that dominate its spectrum: a soft component
(0.2 - 0.6 keV) and a hard component (2 - 10 keV). The hard component
is heavily absorbed and we need to introduce a partial-covering 
absorption in order to match the weak X-ray emission at 0.6 - 2 keV
(see Section~\ref{sec:xray_spec}). The results from our analysis
(Table~\ref{tab:fits}) show that the partial-covering factor could be 
assumed constant but changes of other parameters of the X-ray emission
from \TCRB are noticeable. 

In general between 2017 February and 2018 January, the soft emission 
(0.2 - 0.6 keV) has decreased while the hard emission (2 - 10 keV) has
increased and the flux of the Fe K fluorescent line has increased as
well. 

More specifically, the temperature of the black-body component, 
representative for the soft emission, has increased by $\sim 28$\% 
while its luminosity has decreased by more than an order of magnitude.
On the other hand, the flux of the hard component has increased by a
factor of $\sim 2$ and there are indications that its temperature 
has increased as well, but, we recall that models 
with the same (equal) plasma temperatures of the hot component also
provided very good fits to the observed spectra of \XMM (see
Table~\ref{tab:fits}).

All these changes are an important characteristic of the X-ray
emission from \TCRB during its current active phase and thus provide
valuable pieces of information about the origin of X-rays in this
symbiotic binary. We recall that the boundary layer between the
accretion disk and the white dwarf is the most likely place where the
X-rays form in \TCRB (\citealt{luna_13}; \citealt{luna_18}).
To better understand the physics of this X-ray formation mechanism, 
it is important to see how the characteristics of the X-ray emission 
during the active phase compare with those in a non-active (quiescent) 
phase. For this purpose, the results from our analysis of the \XMM
observations in 2018 and 2017 should be `projected' on those from
X-ray observations taken before the current active phase started.

Our archival search showed that X-ray data suitable for such a 
comparative study, that is, providing spectra in the 0.2 - 10 keV
energy range, have been obtained with \Suzaku and \Swift 
observatories. There is one \Suzaku observation taken in 2006 
September, that is, before the current active phase. On the other 
hand, there are numerous \Swift observations carried out both in the
quiescent phase (2005 June - 2015 September) as well as during the 
active phase (2017 January - 2018 December). 
To derive the global characteristics of the X-ray
emission from \TCRBE, we considered the spectra from the \Suzaku 
observation and two total \Swift spectra, each of the latter 
representative correspondingly of the X-ray emission before and after 
the beginning of the active phase in 2014-2015.
The data reduction and spectral analysis are given 
in Appendix~\ref{app}.

One very important global characteristic of the X-ray emission from
\TCRB is that while the heavily absorbed hard component (2 - 10 keV)
is always present, the strong soft component (0.2 - 0.6 keV) is 
present {\it only} in the active phase.

The appearance of the strong soft component was first reported in the 
analysis of the \XMM observation of 2017 Febuary \citep{luna_18}. In 
addition, we note that this component was not present in the \Suzaku 
and \Swift data, taken before the active phase, while signs of this 
component were found in all of the \Swift observations during the 
active phase (see also Fig.~\ref{fig:spec_suz_swi}). Unfortunately,
there are no \Swift observations (or of another X-ray observatory) 
in the period 2015 October - 2016 December, thus, the earliest time 
the soft component was detected is on 2017 January 18 
(\Swift ObsID 00045776005).

Interestingly, although the hard component is always seen in the X-ray
spectrum of \TCRBE, its strength is quite different in the quiescent
and active phases, namely, it is much weaker in the latter.
This is well illustrated by the observed flux of \TCRB in the (1.5 -
10 keV) energy band, for there is no contribution from the soft 
component at these energies. As seen from Fig.~\ref{fig:lc_suz_swi} 
(left panel), the hard-component flux in the quiescent phase is 
at least one order of magnitude larger than it is in the active
phase. This is also the case with the intrinsic flux of the hard
component in a broad energy range, 0.2 - 80 keV, that is
representative for the hard-component luminosity 
(Fig.~\ref{fig:lc_suz_swi}, right panel). We note that the typical
(average) temperature of the hard component in the quiescent phase is 
kT$ = 16.23$ keV (see Table~\ref{tab:suz_swi}), so, it is considerably 
higher (by a factor of $\approx 2$) than it is in the active phase 
(see Models C  and D in Table~\ref{tab:fits}).
So, there is a clear trend: the stronger the hard component, the 
hotter the plasma emitting the hard X-rays from \TCRBE.

In order to explain (at least qualitatively) all these global 
properties of the X-ray emission
from \TCRB in the quiescent as well as in the active phase, the
following physical picture seems reasonable, we believe.

On a global scale, as discussed by \citet{luna_18}  the mechanism of 
activity in \TCRB could be similar to that suggested for the recurrent 
nova RS Oph. 
Namely, the nova outburst is a result from thermonuclear burning on
the surface of the white dwarf (WD). The necessary `fuel' is deposited 
from the companion star through the accretion disk (AD) around the WD.  
Instabilities in the AD build up the necessary amount of gas through 
numerous phases of activity (\citealt{wynn_08}; \citealt{nelson_11};
\citealt{bollimpalli_18}). And, we witness an active phase in \TCRB 
right now. 

Since \TCRB is a symbiotic star of the $\delta$ class sources of X-ray
emission, it is suggested  that its X-rays originate from the accretion 
disk boundary layer (ADBL) in this white dwarf $+$ M-giant binary 
system (\citealt{luna_13}; \citealt{luna_18}).

We think it is reasonable to assume that the accretion disk is in 
steady-state conditions, when \TCRB is in quiescent phase, so, the 
ADBL produces hard and strong X-ray emission. We note that the gas
velocity in a Keplerian orbit close to the white dwarf in \TCRB is
$\geq 5000$\kms (M$_{WD} = 1.2$\sunmass, \citealt{bel_mik_98}; 
R$_{WD} = 5000$ km as assumed in \citealt{luna_18}), so, high 
temperatures of $> 10$ keV are  potentially possible (e.g., via shocks 
or viscous heating).

When instabilities develop, the ADBL transforms and some considerable 
amount of AD material is being delivered onto the WD surface, thus,
the active phase is switched on. However, it may take some time all
the observational features of an active phase to develop: e.g., the
optical brightening started yet in 2014 \citep{munari_16},  the hard 
X-ray component was still strong in 2015 September but at the same 
time the soft component was not present (evident from the \Swift 
observations, this study; and from the \NuSTAR observation,
\citealt{luna_19}).

The supplied material is located near the AD equator and may be 
gradually diffuses to the WD equator if the WD is inclined with 
respect to the plane of the binary orbit. This on-the-surface material 
is dense and its emission resembles that of a black body. Thus, the 
soft component appears in the X-ray spectrum of \TCRBE. In such a 
case, the 6000-6500-s periodic variability of the soft component 
emission (see Section~\ref{sec:xray_lc} and Fig.~\ref{fig:period}) 
should correspond to the WD rotational period. So, the rotational 
velocity at the equator of the WD in \TCRB  should be $\approx 5$\kms 
for the adopted WD radius (see above) - a value that is not atypical 
for single WDs (e.g., \citealt{karl_05}; \citealt{berger_05}).

Since instabilities have delivered quite a bit of material onto the 
WD surface, some of the ADBL material should have been thus `evacuated' 
and as a result the hard component in the X-ray spectrum of
\TCRB becomes weaker in the active phase. This may explain the almost 
one-order of magnitude difference in the effective accretion rate 
between the quiescent and active phases (see Fig~\ref{fig:lc_suz_swi}, 
right panel). We note that the effective accretion rate was estimated 
in the standard steady-state AD picture, namely, assuming that one 
half of the total accretion energy is released by the AD itself and 
one half of it is released in the ADBL 
(L$_{acc} = G\dot{M}M_{WD}/R_{WD}$, where L$_{acc}$ is the total 
accretion luminosity, $\dot{M}$ is the accretion rate and G is the 
gravitational constant). We have to keep in mind that changes in the
effective accretion rate onto the WD surface do not imply changes of
the global accretion rate from the donor M-giant star. The former
indicate the conditions only in the inner part of an AD, that is, of
its boundary layer, which could have been affected by development of
instabilities.

During the AD evolution back to its steady-state conditions, the ADBL
is gradually building up, thus, its emission (the hard component) is
becoming stronger and it will eventually reach its flux level typical 
for the quiescent phase. 

Interestingly, the ADBL plasma has higher temperature in the quiescent 
phase in comparison with the active phase. The reason could be that 
when instabilities deliver the AD gas onto the WD surface they clear 
off the innermost and hottest part of the ADBL material. As a result, 
the plasma temperature of the hard component lowers in the active 
phase.
We note that this is just a qualitative explanation of this 
temperature difference, so, future observations and numerical AD 
models would be very helpful to get further insights of the physical
processes responsible for it.

On the other hand, no matter whether \TCRB is in quiescent or active
phase the hard component in its spectrum is always highly absorbed. 
This X-ray absorption should be of local origin for its value is more 
than two orders of magnitude above that of the interstellar absorption 
towards \TCRBE. 
Also, analysis of the X-ray emission from \TCRB has lead
to suggestion of the partial-covering high X-ray absorption
(e.g., \citealt{luna_18}). However, the value of partial-covering
factor is very close to unity (see also Table~\ref{tab:fits}) since a
very small amount of X-ray emission is required to 'leak' out in order
to explain the weak emission detected in the 0.6 - 2 keV energy range.
Such a result could be suggestive of some quite special geometry
effects.

We underline that the high X-ray absorption in \TCRB with
a partial-covering factor being {\it extremely} close to unity is
difficult to comprehend. \citet{luna_18} discussed this issue in some
detail and suggested that its resolution could be the presence of a 
clumpy wind from the accretions disk or the physical picture could be
similar to that of the scattering model for the symbiotic binary CH
Cyg \citep{wheatley_06}: the latter requires highly ionized polar
cavity in the AD wind.

On the other hand, if no AD wind is present in \TCRBE, could an
alternative, although speculative, explanation be the following?
Since the hard component is associated with the ADBL 
around the white dwarf in this symbiotic binary and this boundary 
layer likely has small radial extension, the high
X-ray absorption might be due to the accretion disk itself. 
Also, X-rays from the top surface of the ADBL will not be absorbed, 
thus, providing the small `leak-out' emission in the 0.6 - 2 keV
energy range.
In the no-wind case, the soft component (black-body emission forming 
on the WD surface) will not be subject to high X-ray absorption, if 
the AD/orbital inclination is not too close to 90 degrees and the AD 
is indeed geometrically thin: an observer would see the soft component
directly and only the ISM absorption will matter, provided the M-giant
wind is not too massive.

For spherically-symmetric stellar wind with constant velocity and solar
abundances, the hydrogen column density of the M-giant wind along the 
line of sight between the white dwarf in \TCRB and an observer is:
$$
N_{H,wind} = 1.4\times10^{20} \frac{ \dot{M}_9 } {v_{10} \, a_{au}} 
\frac{ \frac{\pi}{2} - \alpha} {\cos \alpha}~,~ cm^{-2}  \nonumber
$$
where $\dot{M}_9$ is the mass-loss rate in units of $10^{-9}$ \sunmass
yr$^{-1}$,
$v_{10}$ is the wind velocity in units ot 10\kms,
$a_{au}$ is the binary separation in astronomical units,
$\alpha$ is defined by
$\cos \alpha = \sqrt{\cos^2 i \sin^2 \omega + \cos^2 \omega}$,
$\omega = 2\pi \phi$, $0 \leq \phi \leq 1$ is the orbital phase 
($\phi = 0$ is the time of maximum radial velocity as in 
\citealt{fekel_00}), $\alpha < 0$ for $\phi > 0.5$,
$i$ is the inclination angle of the binary orbit.

Using the ephemeris from \citet{fekel_00}, we derive that the \XMM 
observations of \TCRB were carried out at orbital phases of 0.954
(2018 January) and  0.455 (2017 February), that is very close to 
binary quadratures.  Given the binary parameters of \TCRB 
($a = 0.9$ au, $i = 60^{\circ}$; see table 2 of \citealt{bel_mik_98}), 
wind veloicity of 10-30\kms and requiring that the M-giant wind 
contributes no more than 10\% of the ISM column density derived from 
analysis of the X-ray spectra of \TCRB (see parameter N$_{H, 1}$ in 
Table~\ref{tab:fits}), we estimate that the mass-loss rate of the 
M-giant stellar wind  should {\it not} be considerably larger than 
$5\times10^{-10}$\sunmass yr$^{-1}$. 

We note that the M-giant wind is an X-ray absorber {\it detached} from
the X-ray formation region (i.e., the ADBL and WD surface) in \TCRBE. 
So, this estimate is valid no matter if both, soft and hard, spectral
components are subject to the partial-covering high absorption or 
only the hard component is.
The \XMM observations thus provide more stringent constraints on the
mass-loss rate of the M giant in \TCRB in comparison with the radio
data that have put an upper limit of 
$8.7\times10^{-9}$\sunmass yr$^{-1}$
(the upper limit from table 2 in \citealt{seaquist_93} was scaled to
the Gaia distnce of 806 pc to \TCRBE).

Finally, it is generally accepted that the Roche lobe overflow of the
M giant is the main source of accretion on the WD in the \TCRB binary.
So, the total accretion rate is not known in advance. If the M giant 
has a stellar wind, some part of it (no more than a half, based on the 
binary parameters) may add to the total accretion rate. The latter 
could in turn be estimated from the X-ray luminosity of the hard X-ray 
component which is associated with the ADBL. 

As seen from Fig.~\ref{fig:lc_suz_swi} (right panel), the accretion 
rate in the quiescent state is not higher than $10^{-9}$\dotM and that 
during the active phase is quite low ($< 10^{-10}$\dotM). So, it does 
not seem feasible that a considerable amount of accreted material will 
be accumulated on the surface of the WD that may lead to a nova 
outbursts with a reccurence period of 80 years (e.g., see tables 2 and 
3 in \citealt{yaron_05}).
But, taking into account that the gravitational energy of the
accreted material also powers the soft component , the situation may
change.  Namely, the accretion rate in active phase could be increased
by a factor of a few  or even boosted cosiderably ($\sim
10^{-7}$\dotM), depending on whether the soft component is subject
{\it only} to the ISM absoorption or is subject to the
partial-covering absorption (see luminosity values
$\log L_{BB}$, $\log L_{BB}^{nc}$, $\log L_{H}$ in 
Table~\ref{tab:fits}).
However, the global spectral fits do not give preferences for either
of these cases: the quality of fits is equal
(see Section~\ref{sec:xray_spec}). So, future X-ray
observations, both in the current active phase as well as in the next
after it quiescent phase, will be very important to get further
details on this important issue.

\section{Conclusions}
\label{sec:conclusions}
In this work, we presented the \XMM data of \TCRBE, obtained during 
its active phase which started in 2014-2015, 
as well as some data taken
with \Suzaku and \Swift in the quiescent phase of this symbiotic 
recurrent nova.  The basic results and conclusions from our analysis of
these data are as follows.

(i)
The \XMM spectra of \TCRB have two prominent components: a soft one
(0.2 - 0.6 keV), well represented by black-body emission, and a 
heavily absorbed hard component (2 - 10 keV), well matched by 
optically-thin plasma emission with high temperature 
(kT $\approx 8$ keV). The soft component was detected for 
the first time in the 2017-February observation \citep{luna_18} and it
is present in the 2018-January observation as well. In addition, a 
weak but statistically significant ($> 5\sigma$ level) emission is 
also present at energies 0.6 - 2 keV.

(ii)
Relatively strong emission-line features are also seen (especially in 
the 2018 data) at high energies: the K-shell fluorescent Fe lines
at $\sim 6.4$~keV (Fe K); the iron He-like triplet at $\sim 6.7$~keV
(Fe XXV); the Fe XXVI L$_\alpha$ at $\sim 6.97$~keV (Fe XXVI) and the
nickel He-like triplet at $\sim 7.8$~keV (Ni XXVII). Presence of such
lines is a sold sign of the thermal origin of the X-rays from \TCRBE.

(iii)
The \XMM observations reveal evolution of the X-ray emission from
\TCRB in its active phase. Namely, the soft component in its spectrum
is decreasing with time (2018 vs 2017) while the opposite is true for
the hard component. 
Comparison with data obtained in the quiescent phase (earlier than
2016) shows that the soft component is typical {\it only} for the
active phase, while the hard component is present in both phases but 
it is considerably stronger in the quiescent phase.

(iv)
Presence of stochastic variability (flickering) on time-scales of 
minutes and hours is confirmed both in X-rays and UV (UVM2 filter of 
the \XMM optical monitor). On the other hand, periodic variability of
6000-6500 s is found for the first time in the soft X-ray emission 
(0.2 - 0.6 keV) from \TCRBE.

(v)
The basic characteristics of the X-ray emission from \TCRB could be
described in the framework of the folowing physical picture. Since
\TCRB is of the class $\delta$ of the X-ray sources among symbiotic
stars,  its heavily absorbed hard component should form in the 
boundary layer between the white dwarf and the accretion disk in this 
symbiotic binary (e.g., \citealt{luna_13}; \citealt{luna_18}). 
Development of instabilities results in delivering a considerable 
amount of the ADBL material onto the surface of the white dwarf in a
relatively short period of time. This marks the beginning of an active 
phase. That material becomes optically thick and thus has black-body 
X-ray emission (the soft component in the X-ray spectrum). So, the 
6000-6500-s periodicity of the soft component is related to the 
rotational period of the white dwarf. During the AD evolution back to
its steady-state conditions, the ADBL is gradually building up, thus,
the hard component in the \TCRB spectrum will eventually return to its
high level of the quiescent phase.
Note that this picture is just an attempt to qualitatively explain the
observational facts about the X-ray emission from \TCRBE. Future X-ray
observations, both in the current active phase as well as in the next
after it quiescent phase, will be very important to get further
insights about the physics of this enigmatic symbiotic recurrent nova.

\section*{Acknowledgements}
This research has made use of data and/or software provided by the
High Energy Astrophysics Science Archive Research Center (HEASARC),
which is a service of the Astrophysics Science Division at NASA/GSFC
and the High Energy Astrophysics Division of the Smithsonian
Astrophysical Observatory. 
This research has made use of the NASA's Astrophysics Data System, and
the SIMBAD astronomical data base, operated by CDS at Strasbourg,
France.
The authors thank an anonymous referee for 
valuable comments and suggestions.

\bibliographystyle{mnras}
\bibliography{tcrb} 

\appendix
\section{X-ray observations of \TCRB with \Suzaku and \Swift}
\label{app}

Here, we briefly describe the data reduction and spectral modelling of
the \Suzaku and \Swift data of \TCRBE.

\begin{figure*}
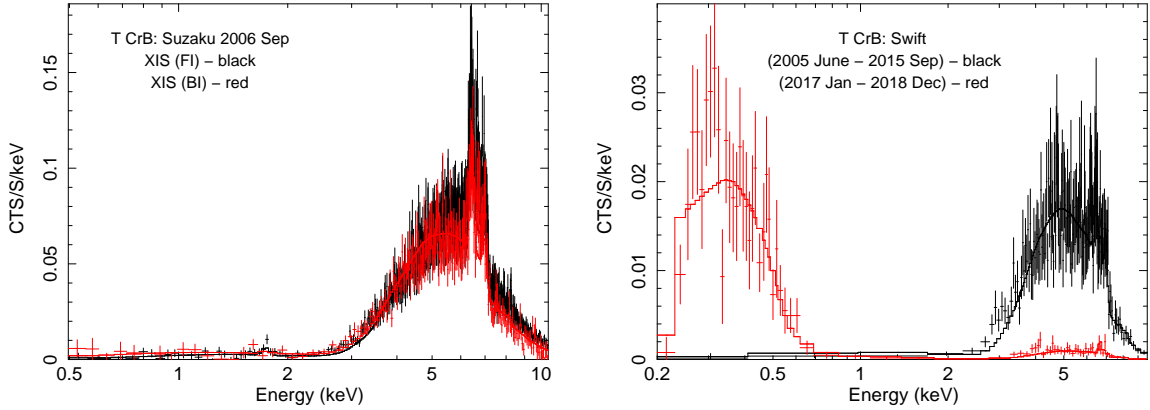

\begin{center}
\includegraphics[width=2.1in, height=3.0in, angle=-90]{figA1a.eps}
\includegraphics[width=2.1in, height=3.0in, angle=-90]{figA1b.eps}
\end{center}
\caption{The background-subtracted spectra of \TCRB from observations
with \Suzaku and \Swift X-ray observatories, overlaid with the model
spectra (Table~\ref{tab:suz_swi}).
XIS (FI) denotes the total spectrum from the three front-illuminated
detectors, while XIS (BI) denotes the spectrum from the
back-illuminated detector on board \SuzakuE.
}
\label{fig:spec_suz_swi}
\end{figure*}

\SuzakuE.
\TCRB was observed by \Suzaku on 2006 September 6 (ObsID 401043010) 
with a nominal exposure of 46.3 ks.
The basic data for our
analysis are from the X-ray Imaging Spectrometers 
(XIS) on-board {\it Suzaku}. We note that one of them (XIS1) is
back-illuminated (BI)  while the other three (XIS0, XIS2, XIS3) are 
front-illuminated (FI) CCDs which allows the extracted spectra from
the latter be combined for further analysis
\footnote{see The 
{\it Suzaku} Data Reduction Guide: 
\url{http://heasarc.gsfc.nasa.gov/docs/suzaku/analysis/abc/}.
Note also that the data were taken berfore the XIS2 loss.}.
We extracted X-ray spectra of \TCRB from the filtered and screened
pipeline events files using the $3\times3$ observational mode which 
resulted in a 38.2 ks effective exposures providing 10847 and 39049 
source counts in the (0.2 - 10 keV) energy range for the BI and total 
FI (XIS0+XIS2+XIS3) spectrum, respectively. 
We adopted 
the current calibration database for {\it Suzaku}, XIS (20181010) and
X-Ray Telescope (XRT; 20110630), to generate the response matrix 
(RMFs) and ancillary files (ARFs).
We note that the \Suzaku telescopes have spatial resolution 
of $\sim 2$~arcmin (expressed as Half-Power Diameter) but
there are no strong X-ray sources in vicinity of \TCRB 
(Fig.~\ref{fig:image}), so, the \Suzaku spectra should be safely
associated with it.

\SwiftE.
\TCRB was observed multiple times in 2005 June 17 - 2018
December 19 (ObsID from 00035171001 to 00035171006; 00081659001 and
00081659002; and from 00045776001 to 00045776039) with typical
exposure time from less than 1 ks to 10 ks.
Following the \Swift XRT Data Reduction
Guide\footnote{\url{http://swift.gsfc.nasa.gov/analysis/xrt_swguide_v1_2.pdf}},
we extracted the source and background spectra for each observation.
Extraction regions had the same shape and size for each data set.
For our analysis, we used the response matrix 
(swxpc0to12s6\_20010101v014.rmf)  provided by
the \Swift calibration
files\footnote{\url{http://heasarc.gsfc.nasa.gov/docs/heasarc/caldb/swift/}}
as of 2019 January,
and we also used the package {\it xrtmkarf} to construct the ancillary 
response file for each data set.

Since we are interested in the global characteristics of the X-ray
emission from \TCRBE, we considered the spectra from the \Suzaku 
observation and two total \Swift spectra, each of the latter 
representative of the X-ray emission in the qiiescent and active 
phase, respectively. In these total \Swift spectra, only
data sets with a minimum of 20 source counts were included, thus, the 
total number of source counts are 3340 (quiescent phase)  and 884
(active phase).  We note that the data of the \Swift 
observations during the active phase (2017 January - 2018 December) 
do not have good photon statistics: only one individual spectrum has 
more than a hundred source counts. As a result, the mean \Swift 
spectrum from observations during the active phase is quite noisy.
For the spectral analysis, the \Suzaku spectra were re-binned to have 
a minimum of 50 counts per bin, while the \Swift spectra were re-bined
with a minimum of 20 (quiescent phase) and 10 (active phase) counts
per bin.

To facilitate the comparison with the results from the \XMM
observations, we used the same analysis approach (modelling) described
in Section~\ref{sec:xray_spec}. For comparison of the global 
characteristics of the X-ray emission from \TCRB in its quiescent and
active phase, we made use of the two-component thermal model
(black-body emission + optically-thin plasma emission).

We fitted simultaneously all three spectra (two from \Suzaku and one 
from \SwiftE) from the observations before the beginning of the active 
phase as they shared the same values of the model parameters that 
determine the spectral shape: the partial-covering absorption, the 
plasma temperature of the hard component. Due to the low quality of the 
mean \Swift spectrum in the active phase (especially, at high
energies), the model parameters for the hard component and the 
partial-covering absorption were kept fixed to their values derived 
from analysis of the \XMM spectra (see Table~\ref{tab:fits}). 
Results from the global fits to the \Suzaku and \Swift spectra are
given in Table~\ref{tab:suz_swi} and Fig.~\ref{fig:spec_suz_swi}.

\begin{table}
\caption{\Suzaku and \Swift Spectral Fits
\label{tab:suz_swi}}
\begin{tabular}{lll}
\hline
\multicolumn{1}{c}{Parameter} & 
\multicolumn{2}{c}{BB + vapec }  \\
\multicolumn{1}{c}{ } &
\multicolumn{1}{c}{quiescent}  & \multicolumn{1}{c}{active} \\
\hline
$\chi^2$/dof  & 1130/1162 & 60/71  \\ 
N$_{H,1}$ (10$^{20}$ cm$^{-2}$)  & 5.18 & 5.18 \\
CF  &
          0.9986$^{+0.0002}_{-0.0002}$ & 0.994 \\
N$_{H,2}$ (10$^{22}$ cm$^{-2}$)  &
          27.5$^{+0.3}_{-0.30}$ & 38.1 \\ 
kT$_{BB}$ (keV) &  & 0.057$^{+0.002}_{-0.002}$ \\ 
R$_{BB}$ (km) &  & 117$^{+71}_{-60}$ \\ 
kT (keV) &
          16.23$^{+0.42}_{-0.44}$ & 7.53 \\
EM ($10^{55}$~cm$^{-3}$) &  21.3$^{+0.2}_{-0.2}$ &
                            1.84$^{+0.14}_{-0.15}$ \\
  &  20.2$^{+0.5}_{-0.5}$ &  \\
%
E$_{l}$ (keV) &
       6.42$^{+0.01}_{-0.01}$ & 6.40 \\
F$_{l}$ ($10^{-6}$ cts cm$^{-2}$ s$^{-1}$) &
       57.2$^{+0.2}_{-0.2}$ & unconstrained \\
F$_{X}$ (0.2 - 10 keV)  &
           \,\,15.8  & \,\,1.07  \\
                                              &
            (59.7) &  (133) \\
      &    \,\,15.0  &           \\
      &     (56.7) &  (133) \\
F$_{X}$ (0.2 - 0.6 keV)  &
           \,\,0.003  & \,\,0.16  \\
                                              &
            (5.1) &  (127) \\
      &    \,\,0.003  &           \\
      &     (4.9) &        \\
\hline

\end{tabular}

{\it Note}.
Results from fits to the \Suzaku and \Swift spectra of \TCRBE.
The adopted \xspec model is: 
$wabs((partcov*wabs)(bbodyrad+vapec) + gaussian)$. 
The labels `quiescent' and `active' 
mark results for the spectra obtained
before 2016 (\Suzaku and \SwiftE) and after 2016 (\SwiftE),
respectively.
All abundances have the same values as in the fits to the \XMM
spectra.  For description of tabulated quantities see Notes to
Table~\ref{tab:fits}.
All fluxes are in units of $10^{-12}$ erg cm$^{-2}$ s$^{-1}$.
For the flux values in column `quiescent', 
the first pair of rows gives the \Suzaku results and the second
pair of rows gives the \Swift results.
Errors are the $1\sigma$ values from the fits.

\end{table}

\bsp    
\label{lastpage}
\end{document}